\documentclass[twocolumn]{aastex63}

\newcommand{\hi}{H {\sc i}~}

\shorttitle{Low Surface Brightness Nebula}
\shortauthors{Zhang et al.}

\begin{document}

%%%%%%%%%%%%%%%%%%%%%%%%%%%%%%%%%%%%%%%%%%%%%%%%
\title{A Mysterious Ring in Dark Space?}

\correspondingauthor{Wei Zhang}
\email{xtwfn@bao.ac.cn}

%\author[0000-0002-1783-957X]{Wei Zhang}
\author{Wei Zhang}
\affiliation{Key Lab of Optical Astronomy, National Astronomical Observatories, Chinese Academy of Sciences, Beijing 100101, China}

\author{Fan Yang}
\affiliation{Key Lab of Optical Astronomy, National Astronomical Observatories, Chinese Academy of Sciences, Beijing 100101, China}

\author{Hong Wu}
\affiliation{Key Lab of Optical Astronomy, National Astronomical Observatories, Chinese Academy of Sciences, Beijing 100101, China}

\author{Chaojian Wu}
\affiliation{Key Lab of Optical Astronomy, National Astronomical Observatories, Chinese Academy of Sciences, Beijing 100101, China}

\author{Hu Zou}
\affiliation{Key Lab of Optical Astronomy, National Astronomical Observatories, Chinese Academy of Sciences, Beijing 100101, China}

\author{Tianmeng Zhang}
\affiliation{Key Lab of Optical Astronomy, National Astronomical Observatories, Chinese Academy of Sciences, Beijing 100101, China}

\author{Xu Zhou} 
\affiliation{Key Lab of Optical Astronomy, National Astronomical Observatories, Chinese Academy of Sciences, Beijing 100101, China}

\author{Fengjie Lei} 
\affiliation{Key Lab of Optical Astronomy, National Astronomical Observatories, Chinese Academy of Sciences, Beijing 100101, China}

\author{Junjie Jin}
\affiliation{Department of Astronomy, School of Physics, Peking University, Beijing 100871, China}

\author{Zhimin Zhou} 
\affiliation{Key Lab of Optical Astronomy, National Astronomical Observatories, Chinese Academy of Sciences, Beijing 100101, China}

\author{Jundan Nie}
\affiliation{Key Lab of Optical Astronomy, National Astronomical Observatories, Chinese Academy of Sciences, Beijing 100101, China}

\author{Jun Ma}
\affiliation{Key Lab of Optical Astronomy, National Astronomical Observatories, Chinese Academy of Sciences, Beijing 100101, China}

\author{Jiali Wang}
\affiliation{Key Lab of Optical Astronomy, National Astronomical Observatories, Chinese Academy of Sciences, Beijing 100101, China}
 
%%%%%%%%%%%%%%%%%%%%%%%%%%%%%%%%%%%%%%%%%%%%%%%%
    
\label{firstpage}

\begin{abstract}

\noindent We report the discovery of a low-surface-brightness (27.42 $\rm
mag~arcsec^{-2}$ in $\textit{g}$ band) nebula, which has a ring-like shape in
the Beijing-Arizona Sky Survey (BASS). Positive detections have been found in
multiband data from far ultraviolet to far infrared, except the $\textit{z}$
band from BASS and W1, W2 from the Wide-field Infrared Survey Explorer. The reddening of the nebula
$E(B - V) \sim$ 0.02 mag is estimated from Infrared Astronomical Satellite (IRAS) 100 $\mu$m intensity and H {\sc
i} column density. With the help of the 3D reddening map from Pan-STARRS 1,
the Two Micron All Sky Survey, and Gaia, the distance to the nebula of about 500 pc from Earth is derived.
Such a low-surface-brightness nebula whose energy can be interpreted by the
diffuse Galactic light could account for the optical counterpart of the
infrared cirrus, which was detected by IRAS more than 30 yr ago. The
ring-like structure might be the ultimate phase of an evolved planetary nebula,
while the central white dwarf star has been ejected from the nebula for an
unclear reason. On the other hand, the ring structure being a superposition of
two close filaments might be another reasonable explanation. Considering the
lack of spectroscopic data and uncertainty in the distance measurement, these
interpretations need to be checked by future observations.

\end{abstract}
    
\keywords{Diffuse nebulae (382); Reflection nebulae (1381); Ring nebulae (1401);
Runaway stars (1417); Neutral hydrogen clouds (1099); Interstellar medium (847); Planetary nebulae (1249);
Interstellar dust extinction (837); Dark interstellar clouds (352);}

\section{Introduction}

The interstellar medium (ISM) is responsible for forming the stars, and is one
of the most important components of galaxies. About 10\% of the baryons in the
Milky Way are found in the ISM. The ISM often appears in a variety of sizes and
structures, which are shaped like of filaments, loops, arcs, sheets, rings, and
expanding shells. The structures are usually related to the ongoing star
formation in the H {\sc ii} regions, supernova expansion, or planetary nebulae (PNs).
Besides, they are also subject to Rayleigh-Taylor instability or even global
reorganizations of large volumes in space. 

The optical diffuse Galactic light (DGL), sometimes called as ``optical cirrus",
was firstly found near the star-forming regions \citep{Elvey-Roach-37}, and then
observed along the whole Galactic plane and even in some high Galactic latitude
areas \citep{Sandage-76}, which was explained as the reflection and scattering of
the starlight by the diffuse ISM. Diffuse emission from the so-called ``infrared
cirrus" at high and intermediate Galactic latitudes has also been detected in
the Infrared Astronomical Satellite (IRAS) 60 and 100 $\mu$m bands \citep{Low-84}, which have shown good spatial
correlations with the DGL \citep{deVries-LePoole-85, Laureijs-Mattila-Schnur-87, Ienaka-13}. 
The spectra of the DGL is consistent with scattered starlight
and the continuum can be reconstructed by a Galactic radiative transfer model
with relatively few large grains \citep{Brandt-Draine-12}. The DGL is detected in
the UV observations as well \citep{Witt-Friedmann-Sasseen-97, Boissier-15}. Complementing
multiwavelength observations including UV, optical, and IR, the DGL could be a very
helpful tracer to better understand the nature of the Galactic ISM and details
of the star-forming process \citep{Miville-Deschenes-16}.

The current generation of deep optical surveys allow us to detect faint galaxy
satellites \citep{McConnachie-12} and explore galactic substructures as tidal
features and stellar streams \citep{Ibata-Mouhcine-Rejkuba-09, Tanaka-11,
Ibata-14, Martin-14, McConnachie-18}. The DGL can be a significant contaminant
for extragalactic studies, even at high Galactic latitudes, particularly when
the focus is low-surface-brightness galaxies (LSBGs), or diffuse stellar halos
around massive galaxies. The contaminant becomes more prominent when the
surface brightness is fainter than 26 $\rm mag~arcsec^{-2}$ \citep{Duc-15,
Greco-18}.  The presence of the diffuse light scattered by Galactic dust clouds
could significantly bias our interpretation of low-surface-brightness features
and diffuse light observed around galaxies and in the clusters of galaxies
\citep{Cortese-10}.  For example, \cite{Mihos-05} first find a
low-surface-brightness plume near the interacting pair NGC 4435/4438 in the
Virgo cluster. This feature has been interpreted as a tidal stream, until it
was found more like Galactic cirrus by \cite{Cortese-10} using multiband data
including far-UV (FUV), far-IR (FIR), H {\sc i}  and CO emissions.  Studied in
the 1960s, Arp's loop has been thought to be a tidal tail, material pulled out
of M81 by gravitational interaction with its large neighboring galaxy M82
\citep{Arp-65}.  But a recent investigation demonstrates that much of Arp's
loop likely lies within our own Galaxy \citep{Sollima-10}.
\cite{Roman-Trujillo-Montes-19} systematically searched the Galactic cirrus
based on deep optical photometry in the Sloan Digital Sky Survey (SDSS)
Stripe82 region.  The results showed that the higher 100 $\mu$m emission, the
redder color is for the Galactic cirrus. And in most cases, the optical colors
of the Galactic cirrus differ significantly from those of extragalactic
sources. 

During the process of searching LSBGs in the Beijing-Arizona Sky Survey (BASS),
we find a low-surface-brightness nebula with a ring-like shape whose
$\textit{g}$ band surface brightness is down to 27.42 $\rm mag~arcsec^{-2}$.
Observation data suggest that as the wide-field photometric surveys go deeper,
more nebulae at the faint end of the DGL similar to this one are likely to be
found. The symmetrical shape can easily raise a question: could this
low-surface-brightness nebula be a supernova remnant or a PN? Through detailed
analysis of multiband data from FUV to 21 cm, the possibility that the circular
shape of the nebula is the ultimate phase of a PN that exploded about 255,000
yr ago is discussed. 

The paper is structured as follows. In \S2 we introduce the multiband
observations from UV to H {\sc i}. We calculate the extinction and distance of
the nebula in \S3. We discuss several possible mechanisms of the
formation of the ring-like nebula in \S4. Finally we end the paper with the conclusion
and discussion in \S5.

\begin{figure}%[ht]
\centering
\includegraphics[width=\linewidth]{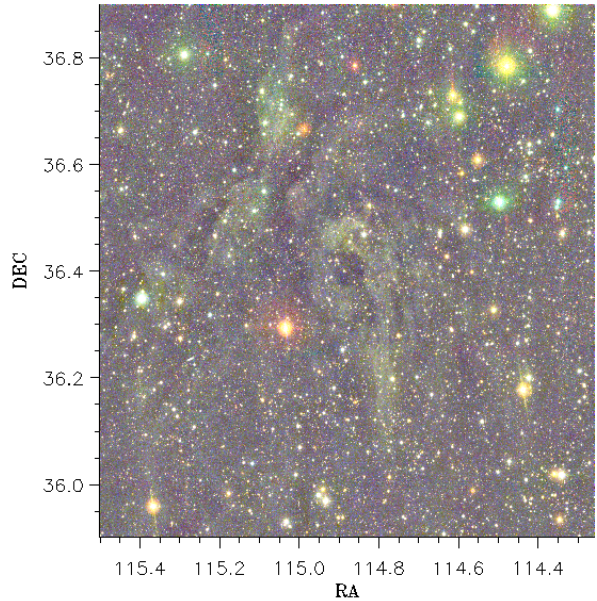}
\caption{RGB color image of the newly discovered low-surface-brightness nebula with a FOV of 1$^{\circ} \times$ 1$^{\circ}$. Three optical bands of smoothed images including $\textit{u}$, $\textit{g}$ and $\textit{r}$ are stacked together to generate the colored image. The ring-like shape in the center and the long tail extending to the south form a shape like the number 9.}
\label{fig:color_img}
\end{figure}

\begin{figure*}
\centering
\includegraphics[width=\textwidth]{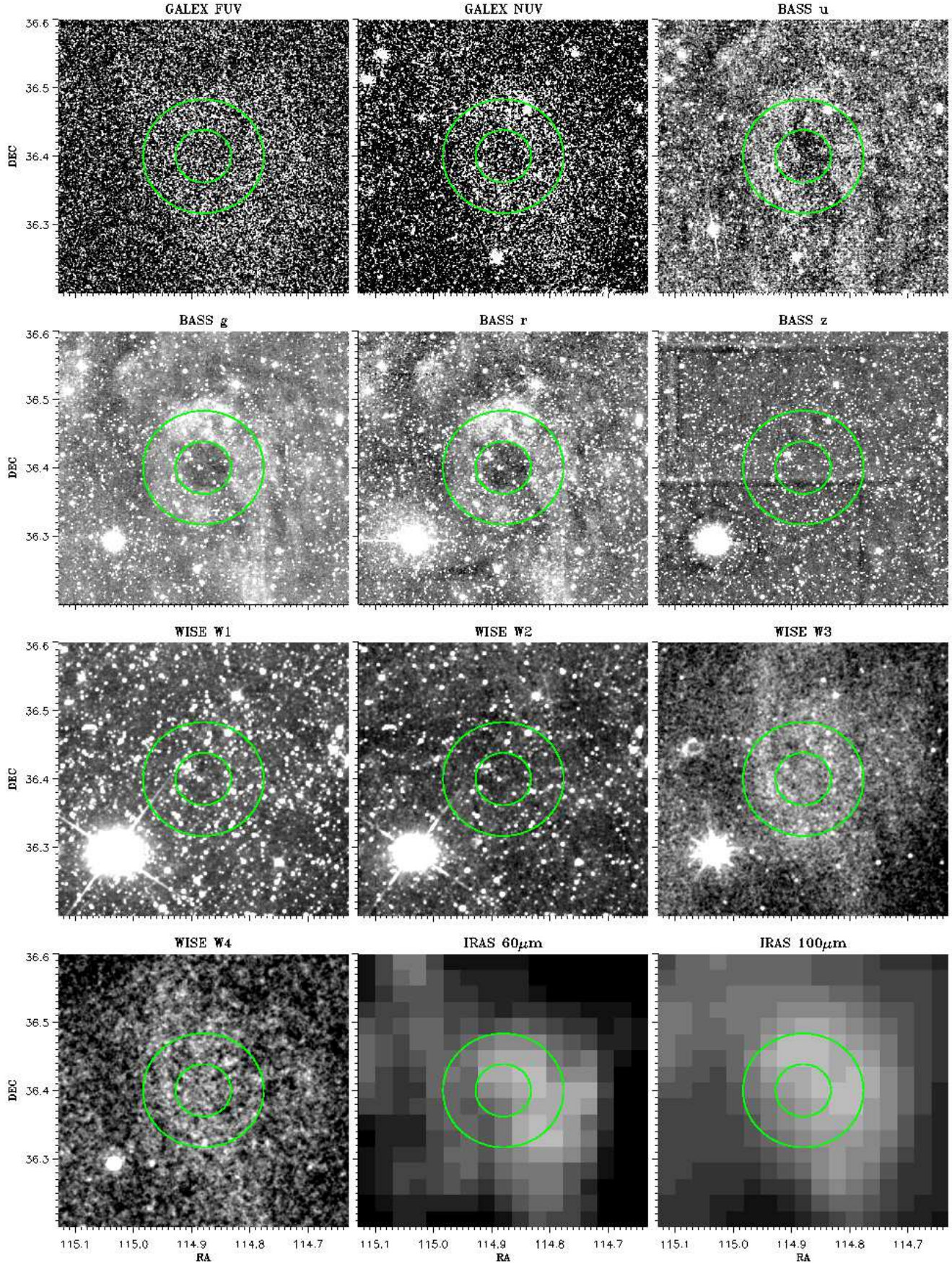}
 \caption{Multiband observations of the target area with a FOV of $24'\times24'$. Top row: GALEX FUV, GALEX NUV, and  BASS $\textit{u}$. Second row: BASS $\textit{g}$, BASS $\textit{r}$, and BASS $\textit{z}$. Third row: WISE W1, WISE W2, and WISE W3. Bottom row: WISE W4, IRAS 60$\mu$m, and IRAS $100\mu$m. All bands except $\textit{z}$, W1, and W2 have shown positive detections of the nebula structure. The two green circles are $4\farcm6$ and $10'$ in diameter, respectively.}
\label{fig:12bands}
\end{figure*}

\begin{figure}
\centering
\includegraphics[width=\linewidth]{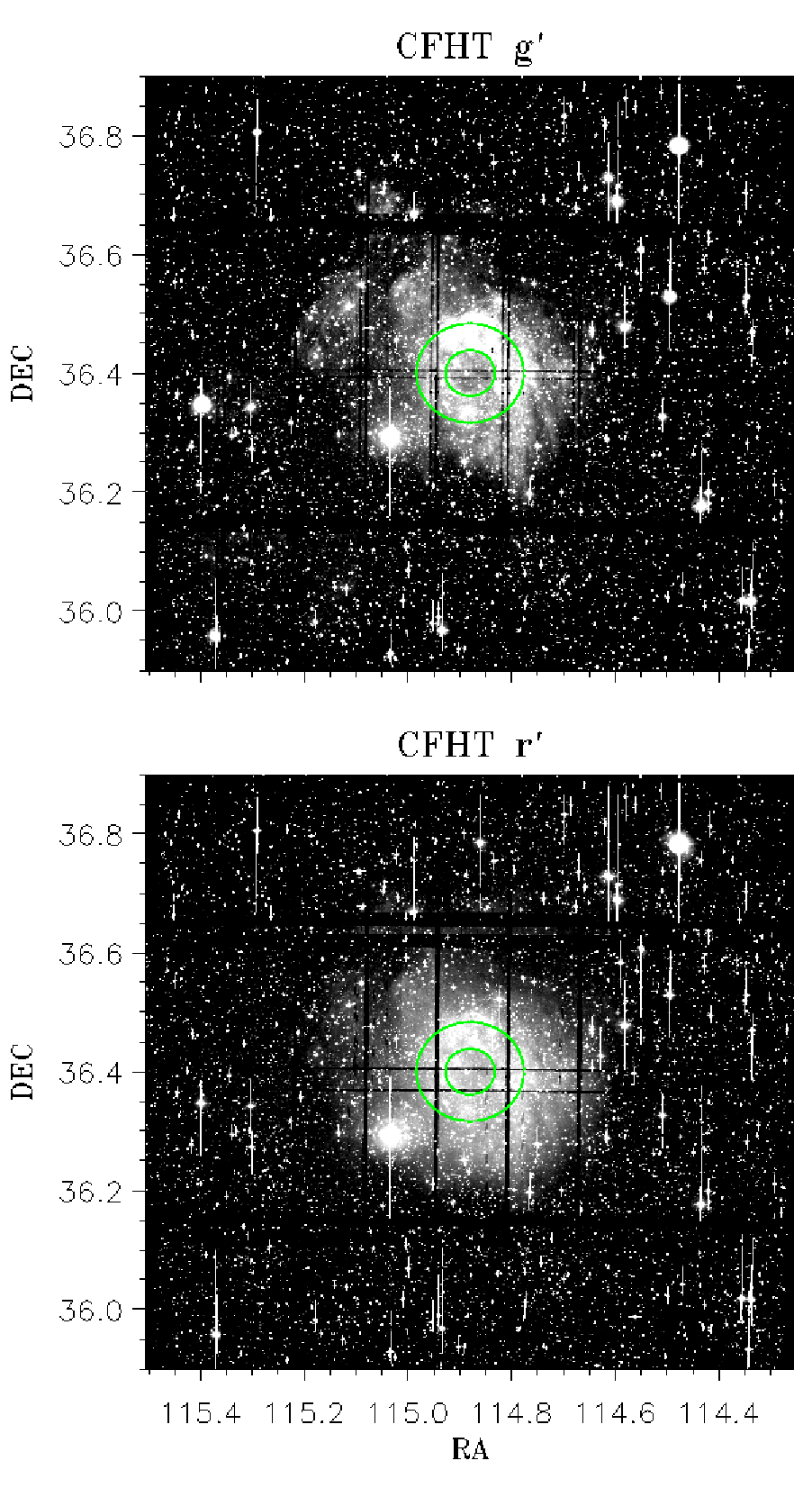}
 \caption{Images in the $\textit{g'}$ and $\textit{r'}$ bands observed by CFHT with a FOV of 1$^{\circ} \times$ 1$^{\circ}$. Similar to Figure \ref{fig:12bands}, the two green circles are $4\farcm6$ and $10'$ in diameter, respectively.}
\label{fig:cfht}
\end{figure}

\section{Observations}

Our work uses photometric data of the BASS in the $\textit{g}$ and $\textit{r}$
bands with the prime focus of the Bok 2.3 m telescope. The Bok Telescope, owned
and operated by the University of Arizona, is situated on Kitt Peak and is
adjacent to the 4 m Mayall Telescope. The 90Prime instrument is a prime focus
8k $\times$ 8k CCD imager, with four University of Arizona Imaging Technology
Laboratory (ITL) 4k $\times$ 4k CCDs that have been thinned and UV optimized
with a peak quantum efficiency of 95\% at 4000~\AA.  Since BASS only contains
$\textit{g}$ and $\textit{r}$ filters, the $\textit{u}$ and $\textit{z}$ band
images were taken specifically using the Bok Telescope and the Mayall
Telescope, so as to cover the whole optical wavelength range.  For the
$\textit{u}$, $\textit{g}$, $\textit{r}$, and $\textit{z}$ bands, the total
integration times are about 3750, 750, 580, and 280 s; the image resolutions
(FWHMs), are 1$\farcs$3, 1$\farcs$7, 1$\farcs$3  and 1$\farcs$3; and the
magnitude limits at the 5$\sigma$ level for point sources can reach 24.5, 24.2,
23.6, and 23.0 mag in AB systems, respectively \citep{Zou-19}. The target field
is centered at (RA., decl.)$_{\rm J2000}$ = (114$\fdg$88, 36$\fdg$40) and it is
about 1$\fdg$4 $\times$ 1$\fdg$1 wide.  Three optical images including
$\textit{u}$, $\textit{g}$ and $\textit{r}$ are smoothed and stacked to
generate a color image with a field of view (FOV) of 1$^{\circ} \times$
1$^{\circ}$ as shown in Figure \ref{fig:color_img}, in which a ring-like nebula
is spotted.  More interesting is that this ring and a spur extending to the
south form a shape like the number 9.

In order to eliminate the possibility of a ghost image or artifacts, we compare
the images from the Galaxy Evolution Explorer (GALEX), Wide-field Infrared
Survey Explorer (WISE) and IRAS. As shown in Figure \ref{fig:12bands}, four
optical bands from the BASS observations, near-UV (NUV) and far-UV (FUV) from
GALEX, W1, W2, W3, and W4 from WISE, and 60 and 100 $\mu$m from IRAS are
centered at the same frame. All bands except $\textit{z}$ from BASS and W1 and
W2 from WISE have shown positive detections. In all images, two green circles
with 4$\farcm$6 and $10'$ in diameter are plotted over the nebula.

\subsection{BASS Data}

Normally, a nebula with a mean surface brightness as faint as 26
$\rm mag~arcsec^{-2}$ would be neglected and subtracted simply as a background when
doing sky subtraction. As more deep optical surveys are to be carried out, the
possibilities of finding these faint objects are increasing. Since
the nebula itself is only several ADUs higher than the background, it is hard
to pick out by the naked eye. To enhance the image of the nebula, a Gaussian
smooth kernel with a radius of four pixels is applied. After being smoothed, the
mysterious ring-like feature has gradually emerged in the $\textit{u}$,
$\textit{g}$, $\textit{r}$ bands of the BASS survey. Note that the ring
shape in the $\textit{u}$ band is less obvious than that in the $\textit{g}$ or
$\textit{r}$ bands, which might be caused by the lower signal-to-noise ratio.

Since the central region of the nebula has a ring-like structure, with an inner
radius of $2\farcm3$ and an outer radius of 5$'$, we mainly focus on this part
and try to reconstruct the spectral energy distribution of this region to
better understand its property. To precisely measure the flux of the selected
central region of the nebula in optical bands, we use SExtractor to detect and
mask sources such as stars and galaxies.  As the whole area is covered by the
nebula, this brings difficulties in defining the true sky and having it
subtracted. We choose a nearby area which has the central coordinate (RA.,
decl.)$_{\rm J2000}$=(114$\fdg$4, $36\fdg40$), with a radius of 1$\farcm$8, as the
background. (Note that this region is also taken as the background for all the
other bands). The median value of the flux in this area is employed as the sky
value. The flux of the masked areas is also replaced by the mean flux of the
cloud.  The AB magnitude is $m = -2.5log(f)+zp$, where $f$ is the corrected
flux of the central region and $zp = 30$, given by BASS. Finally we derive the
apparent magnitudes $\textit{u}$ = 15.28, $\textit{g}$ = 13.55, $\textit{r}$ =
13.36 and surface brightnesses $\rm SB_\textit{u}$ = 28.65, $\rm SB_\textit{g}$
= 27.03, $\rm SB_\textit{r}$ = 26.73 $\rm mag~arcsec^{-2}$. Note that optical
color of the nebula is $\textit{g}-\textit{r}$ = 0.31, corresponding to
stars with an effective temperature of 6100 K \citep{Ramrez-Melendez-05}.  

\subsection{CFHT Data}

Observations have also been carried out at the 3.6 m Canada-France-Hawaii Telescope
(CFHT), with the wide-field imaging MegaCam camera, covering a 1 deg$^2$ area in
the sky. MegaCam is made of 36 CCDs with gaps of $13\arcsec-80\arcsec$ between them.
The images in the $\textit{g'}$ and $\textit{r'}$ bands were obtained on 2018 January 17, 18,
and 20, with a total integration time of $\sim$ 1.5 hr for each band.
CFHT images with a FOV of $1^{\circ} \times 1^{\circ}$ shown in Figure
\ref{fig:cfht} have been used to check whether there are fainter substructures
undetected by BASS. The ring-like structure is confirmed by the CFHT data.
Besides, compared to BASS, CFHT detected even fainter optical emissions in the
outer region and the central hole region, which makes the ring shape 
less obvious than that in the BASS $\textit{g}$ or $\textit{r}$ data. On the other
hand, the fainter structure detected by CFHT indicates that CFHT has an even lower
detection limit than BASS, because of a larger mirror and a longer exposure
time.

\subsection{UV Data}

Images in the FUV and NUV bands observed by GALEX \citep{Martin-05} were
downloaded from the Mikulski Archive for Space Telescopes
website\footnote{https://archive.stsci.edu/missions-and-data/galex-1/}. The FUV
and NUV image resolutions, FWHMs, are 4$\farcs$5 and 6$\farcs$0, respectively.
The exposure time is 215 s for both FUV and NUV.  The intensity in the
central region is lower than that in the ring region, but no circular hole is
found in these two bands, which might be caused by the lower signal-to-noise
ratio. These images are used to derive the counts per second (CPS) of the
chosen area.  A Gaussian kernel with a radius of four pixels is applied to
smooth the image before our using SExtractor to generate the mask of point
sources. We then replace the CPS of the pixels in all masked areas with the
median value of the chosen area.  The background is subtracted using the same
method as for the optical data. The AB magnitudes of FUV and NUV for the nebula
in the ring region can be converted from the CPS \citep{Morrissey-07}:

\begin{eqnarray}
\textup{FUV AB magnitude} = -2.5\textup{log}\left ( \textup{CPS} \right )+18.82\\
\textup{NUV AB magnitude} = -2.5\textup{log}\left ( \textup{CPS} \right )+20.02
\end{eqnarray}

The magnitudes of FUV and NUV are FUV=14.82 and NUV=14.57, respectively. 
The mean FUV flux intensity of the nebula is measured as $\rm S_{FUV}\sim1.05\times10^{-8}~erg~s^{-1}
~cm^{-2}~\AA^{-1}~sr^{-1}$.

\subsection{IR Data}

The WISE \citep{Wright-10} W1, W2, W3, and W4 images (with FWHMs of
6$\farcs$1, 6$\farcs$4, 6$\farcs$5, and 12$\arcsec$, respectively), and the
Infrared Astronomical Satellite \citep[IRAS,][]{Neugebauer-84} 60 $\mu$m and
100 $\mu$m images (with FWHMs of 4$\farcs$0 and 4$\farcs$3, respectively) were
downloaded from the Infrared Science Archive
website\footnote{https://irsa.ipac.caltech.edu/frontpage/}. From Figure \ref{fig:12bands},
we can see that there are positive detections in the W3 and W4 bands, while no
emission is found in the W1 or W2 band. The visible emission at the W3 and W4
images indicates thermal fluctuations of very small grains and/or polycyclic
aromatic hydrocarbons (PAHs) within this nebula
\citep{Szomoru-Guhathakurta-98}. Similar to the CFHT, the structure in the W3
and W4 data is more extended than that in the BASS $\textit{u}$, $\textit{g}$,
and $\textit{r}$ images.  Besides, no circular hole is found in the center of
the structure.

The signals are also significant in the IRAS 60 and 100 $\mu$m images. At
first glance, the structure has an offset in these two bands, compared to that
in the BASS optical bands. In fact, this is because the intensity in the northwest part
is obviously higher than that in the southeast part, which will be analyzed in
\S4.4. Besides, the spur to the south is clearly detected in these two bands.
The mean brightness of 100 $\mu$m, $S_{100(tot)}$, in the ring region of the
nebula is 5.10 $\rm MJy~sr^{-1}$. After removing the background using the same way as
described in \S2.1, the mean brightnesses of 60 and 100 $\mu$m, $S_{60(neb)}$ and
$S_{100(neb)}$, in the ring region of the nebula are about 0.50 and 2.34
$\rm MJy~sr^{-1}$, respectively.  Assuming the emissivity spectral index $\beta=2$,
the dust temperature ($T_d$) is estimated by the observational ratio of $60-100$
$\mu$m:
$R=S_{60(neb)}/S_{100(neb)}=0.6^{-(3+\beta)}\frac{e^{144/T_d}-1}{e^{240/T_d}-1}$,
resulting in $T_d=22$ K \citep{Schnee-05}.

\begin{figure}%[ht]
\centering
\includegraphics[width=\linewidth]{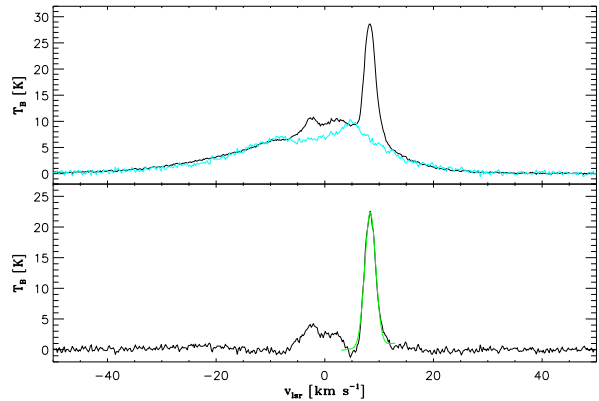}
\caption{The mean H {\sc i} spectrum in the directions of the nebula. The narrow component, contributed by the nebula, has been fitted by a Gaussian profile, to derive $\rm v_{lsr,neb}$, $\rm \sigma(v_{lsr,neb})$ and column density $N_{HI}(neb)$.}
%\label{fig:RV_HI}
\label{fig:N_HI}
\end{figure}

\begin{figure*}%[ht]
\centering
\includegraphics[width=\textwidth]{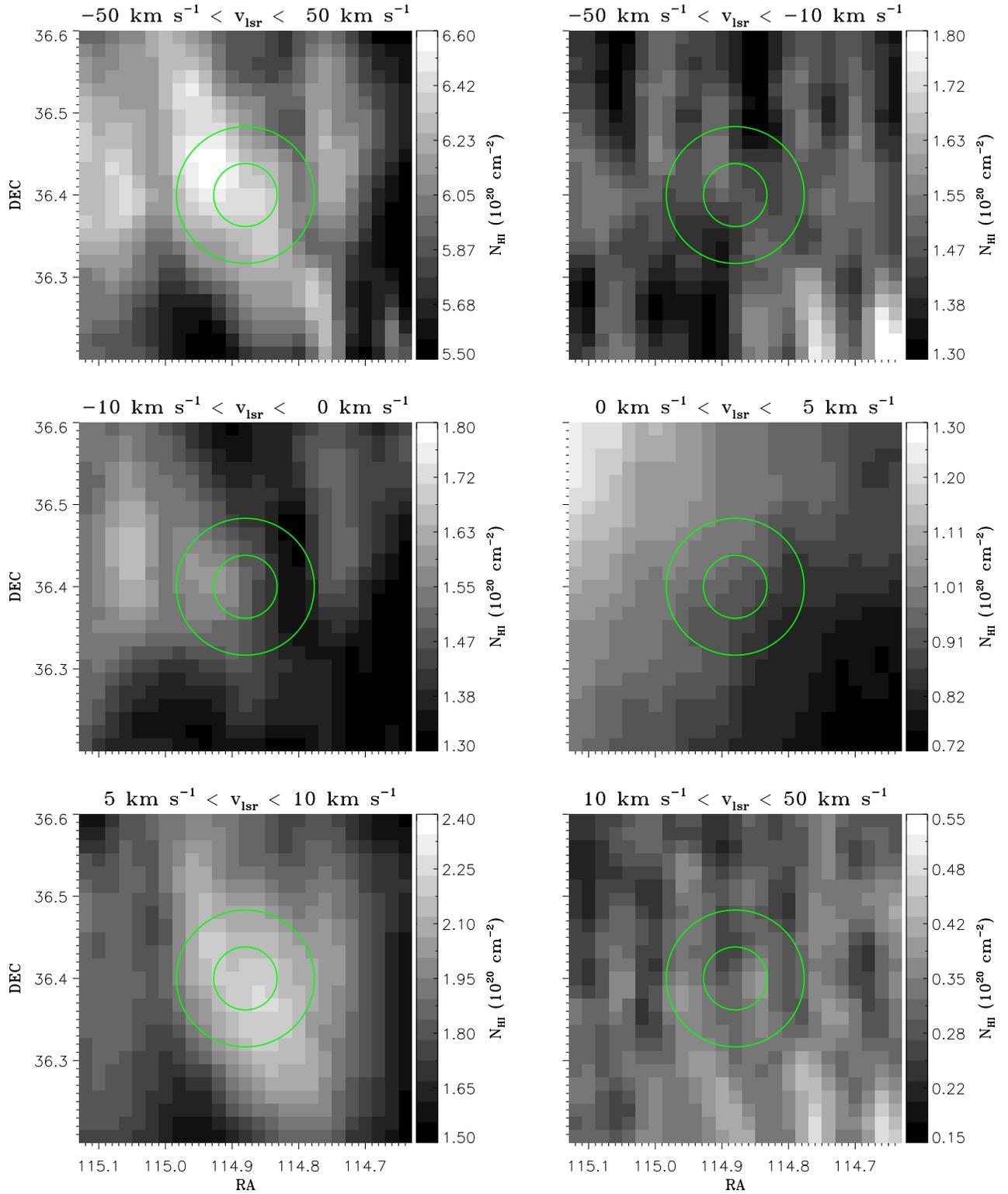}
 \caption{2D spatial distributions of N$_{HI}$ with a FOV of $24'\times24'$. Similar to Figure \ref{fig:12bands}, the two green circles are $4.6'$ and $10'$ in diameter, respectively.}
\label{fig:hi-2d}
\end{figure*}

\subsection{\hi Data}

The H {\sc i} data is derived from the Galactic Arecibo $L$-band Feed Array H {\sc
i} (GALFA-H {\sc i}) Survey Data release
2\footnote{https://purcell.ssl.berkeley.edu}.  The data are served in two sets
of data cubes with $4'$ angular resolution. In this work, we use the ``Narrow"
set cubes with the full resolution of 0.184 $\rm km~s^{-1}$ over the velocity range
$\rm |v_{lsr}|~\le~188.4~km~s^{-1}$.  Applying the optically thin assumption, the
H {\sc i} column density can be obtained by integrating the H {\sc i} spectra
over a velocity interval \citep{Peek-18}:

\begin{equation}
N_{\rm HI} = 1.823 \times 10^{18}{\rm~cm^{-2}} \frac{\int T_B d_\nu }{\rm K~km~s^{-1}}
\label{equ:NHI}
\end{equation}

%H {\sc i} spectra in the direction of the ring region is showed in Figure
%\ref{fig:N_HI}. 

The mean H {\sc i} spectrum in the direction of the ring region is shown
by the black line in the top panel of Figure \ref{fig:N_HI}, and the mean H {\sc i}
spectrum in the direction of the background region is shown by the cyan line in the
same panel. After subtracting the background spectrum, the residual spectrum is
shown in the bottom panel. We can see there are still two components in the
residual H {\sc i} spectrum, the wider and weaker component with velocity
$\rm v_{lsr}\sim$ 0 $\rm km~s^{-1}$ comes from the residual background
(hereafter named as rb) H {\sc i} emission, which is mainly contributed by the
Galactic disk, while the narrower and stronger component comes from the nebula.
The total H {\sc i} column density $N$$_{\rm HI}$(tot) along the line of sight in the
direction of the nebula is 6.28 $\times10^{20}\rm~cm^{-2}$, by integrating the H {\sc
i} spectra over the velocity interval $\rm -50\ km\ s^{-1} < v_{lsr} < 50\ km\
s^{-1}$. The H {\sc i} column density of the residual background
$N_{\rm HI}$(rb) is $\rm 4.21\times10^{19}~cm^{-2}$, by integrating the residual H {\sc
i} spectrum over the velocity interval $\rm -10\ km\ s^{-1} < v_{lsr} < 5\ km\
s^{-1}$. We also fit the narrower component using a simple Gaussian profile
(shown by the green line in the bottom panel in Figure \ref{fig:N_HI}), the 
best-fitting result shows that the velocity of the nebula $\rm v_{\rm lsr,neb}$ is $\sim$
8.36 $\rm km~s^{-1}$ with a dispersion $\rm \sigma(v_{\rm lsr,neb)}$ of 1.02 $\rm km~s^{-1}$
and the H {\sc i} column density of the nebula $N_{\rm HI}$(neb) is about
$\rm 9.97\times10^{19}~cm^{-2}$.

\begin{deluxetable*}{lccccccccccccc}
\tablecaption{Observational properties of the nebula centered at (R.A., decl.)$_{\rm J2000}$~=~(114.88$^{\circ}$, 36.4$^{\circ}$). \label{tbl:sed}}
\tablewidth{0pt}   
\tablehead{
%\colhead{Survey} \vline & \multicolumn{2}{c}{GALEX} \vline  & \multicolumn{3}{c}{BASS} \vline  & \colhead{DzLS} \vline & \multicolumn{4}{c}{WISE} \vline  & \multicolumn{2}{c}{IRAS} \vline & \colhead{HI}\\
\colhead{Survey} & \multicolumn{2}{c}{GALEX} & \multicolumn{4}{c}{BASS} & \multicolumn{4}{c}{WISE} & \multicolumn{2}{c}{IRAS} & \colhead{GALFA-HI}\\
\hline
%\colhead{Band} & \colhead{FUV} & \colhead{NUV} & \colhead{u} & \colhead{g} & \colhead{r} & \colhead{z} & \colhead{W1} & \colhead{W2} & \colhead{W3} & \colhead{W4} & \colhead{60$\mu$m} & \colhead{100$\mu$m} & \colhead{21cm} 
\colhead{Band} \vline & \colhead{FUV} & \colhead{NUV} \vline  & \colhead{$\textit{u}$} & \colhead{$\textit{g}$} & \colhead{$\textit{r}$} & \colhead{$\textit{z}$} \vline & \colhead{W1} & \colhead{W2} & \colhead{W3} & \colhead{W4} \vline & \colhead{60$\mu$m} & \colhead{100$\mu$m} \vline & \colhead{21cm} 
}
%\decimalcolnumbers
\startdata
FWHM   & $4\farcs5$ & $6\farcs0$ & $1\farcs3$ & $1\farcs7$  & $1\farcs3$  & $1\farcs3$ & $6\farcs1$  & $6\farcs4$  & $6\farcs5$  & 12$^{\prime\prime}$  & $4\farcm0$ & $4\farcm3$ & $4\farcm0$ \\
\hline
mag (AB)     & 14.82 & 14.57 & 15.28 & 13.66 & 13.36 & 14.82 & 11.12 & 11.73 & 9.40  & 9.29  &    &    &    \\
mag$^c$(AB) & 15.21 & 14.96 & 15.67 & 14.05 & 13.75 & 15.21 & 11.51 & 12.12 & 9.79  & 9.68  &    &    &    \\
\hline 
SB (mag~arcsec$^{-2}$)      & 28.19 & 27.94 & 28.65 & 27.03 & 26.73 & 28.19 & 24.49 & 25.10 & 22.77 & 22.67 &    &    &    \\
SB$^c$ (mag~arcsec$^{-2}$) & 28.58 & 28.33 & 29.04 & 27.42 & 27.12 & 28.58 & 24.88 & 25.49 & 23.16 & 23.05 &    &    &    \\
\hline 
I$_\nu~(Jy)$   &    &    &    &    &    &   &   &   &    &    & 2.59  & 12.27 &    \\
I$_\nu^c~(Jy)$   &    &    &    &    &    &   &   &   &    &    & 1.82  & 8.59  &    \\
\hline 
S$_\nu~(MJy~sr^{-1}$)        &    &    &    &    &    &   &   &   &    &    & 0.50  & 2.34 &    \\
S$_\nu^c~(MJy~sr^{-1}$)   &    &    &    &    &    &   &   &   &    &    & 0.35  & 1.64  &    \\
\hline 
$N^c_{HI}~(cm^{-2})$   &    &    &    &    &    &   &   &   &    &    &    &    &  9.97$\times10^{19}$  \\
\enddata
\tablecomments{~$^c$ This parameter denotes that the residual background correction has been considered.}
\end{deluxetable*}

The spatial distributions of the H {\sc i} signals with a FOV of
$24^{\prime}\times24^{\prime}$ are shown in Figure \ref{fig:hi-2d}.  The velocity
interval used to integrate the spectrum is shown on the top of each panel. The
structure is clearly found when the velocity ranges from 5 to
$\rm 10\ km\ s^{-1}$. Similar to the CFHT, W3, W4, and IRAS 60 and 100 $\mu$m,
the structure is more extended than that in the BASS $\textit{g}$ and $\textit{r}$ images,
while no ring shape is found, which should be caused by the lower resolution
(FWHM $\sim$ 4$^{\prime}$).

\subsection{Summary of the Multiband Observations}

The ring structure is detected by BASS $\textit{g}$ and $\textit{r}$,
and is confirmed by CFHT $\textit{g'}$ and $\textit{r'}$. Because CFHT has a 
lower detection limit, the nebula and central depression are present, making
the ring shape being less obvious. This nebula is also detected by BASS
$\textit{u}$, GALEX FUV, NUV, WISE W3, W4, IRAS 60, 100 $\mu$m, and GALFA-H {\sc i}
data, while no circular hole is shown in the central region, which might be caused
by the lower signal-to-noise ratio or the lower spatial resolution. We hence declare
that the structure is described as a ring only in optical $\textit{g}$ and
$\textit{r}$ data, and the ring might be found in other bands in future using
higher-quality data. Besides, the intensity of the northwest part of the
structure is obviously higher than the southeast part in IRAS 60 and 100 $\mu$
m images. The different spatial distribution between IRAS 60, 100$\mu$m and
H {\sc i} indicates the dust to gas ratio might not be uniform in this region.

Most of the properties of the nebula are summarized in Table \ref{tbl:sed},
including the image resolution FWHMs, visual magnitudes in AB systems, surface
brightnesses, and intensities. The magnitudes of W3 and W4 were converted from
the WISE Vega system to the AB system, by adding the offsets of 5.242 and 6.604 mag,
respectively. As shown in \S2.5, the components along the same line of sight
can be easily resolved by the H {\sc i} spectra. For other bands, most of the
background can be subtracted by taking a nearby region as the background. But
this background is not identical to that in the targeted region.  For the
nebula in this work, the fraction of the residual background
$f_{rb}=\frac{N_{HI}(rb)}{N_{HI}(rb)+N_{HI}(neb)}$ can be estimated to be 30\%.
We then apply this correction to the other bands, and mark the parameters with
($^c$) in Table \ref{tbl:sed}. This correction makes the magnitude and surface
brightness fainter by about 0.38 mag. In this result, the surface brightness in the
BASS $\textit{g}$ band reaches to 27.42 mag, and the brightness $S^c_{100}$
turns into 1.64 $\rm MJy~sr^{-1}$. Besides, the magnitudes and surface brightnesses
of BASS \textit{z}, W1, and W2 are also given, although the signals in these
bands cannot be recognized by the naked eye.

\section{Reddening and Distance}

\subsection{Reddening}

The column gas density has been found to be well correlated with the
interstellar extinction, hence the $V$ band extinction can be measured by $A_V =
5.3\times10^{-22}N_{\rm HI}$ $\rm atoms~cm^2$
\citep{Bohlin-Savage-Drake-78,Rachford-09}, which yields values of $A_V$(tot) =
0.33 mag and $A_V$(neb) = 0.05 mag.

Following the work of \cite{Savage-Mathis-79}, the 100 $\mu$m brightness
$S_{100}$ can be converted into the $V$ band extinction by $A_V = 0.053
(S_{100}/1 \rm MJy~sr^{-1})$, resulting in the total extinction in the direction of
the nebula $A_V$(tot) = 0.27 mag and the extinction of the nebula alone
$A_V$(neb) = 0.09 mag.

Combining the results from above two methods, the total extinction along the line
of sight is $A_V(tot)$ = 0.30 mag, and the extinction of the nebula alone
is $A_V(neb)$ = 0.07 mag. By adopting $R_V=3.1$, the total reddening value
$E(B - V)_{tot}$ is about 0.097 mag, and the reddening of the nebula alone
$E(B - V)_{neb}$ is about 0.02 mag.

%Topical subheadings are allowed.

\subsection{Distance}

Distance is a basic parameter to do further analyze for this nebula. At first
glance, this ring-like nebula comprises material ejected from an evolved
asymptotic giant branch (AGB) star. If this is true, there should be a white
dwarf star located at the center of the ring. However, we have analyzed the
stars around the center, and found that none of them is hot enough to be a
white dwarf. So there are several possible explanations: (1) the nebula is not a
relic of an evolved PN, (2) the white dwarf star becomes too
faint to be detected by the BASS photometry, or (3) the white dwarf star has run
away from the center. Whatever the truth is, it is impossible to derive the
distance of the system from the parallax of the central star.

In principle, the kinematic distance can be obtained from the radial velocity
of the gas by adopting a specific Galactic rotation curve and solar motion
parameters. Besides the two traditional methods, \cite{Wenger-18} also
developed a Monte Carlo method to derive the kinematic distances, in which the
probability distribution functions of the kinematic distances are determined by
resampling all measured and derived parameters within their uncertainties.
Unfortunately, the kinematic distances are very inaccurate toward the Galactic
Center and the Galactic Anticenter due to velocity crowding. So we did not try
to derive the kinematic distance to the nebula centered at
$(l,~b)=(183\fdg09,~24\fdg82)$, because it is in the direction of the Galactic
Anticenter.

\begin{figure}%[ht]
\centering
\includegraphics[width=\linewidth]{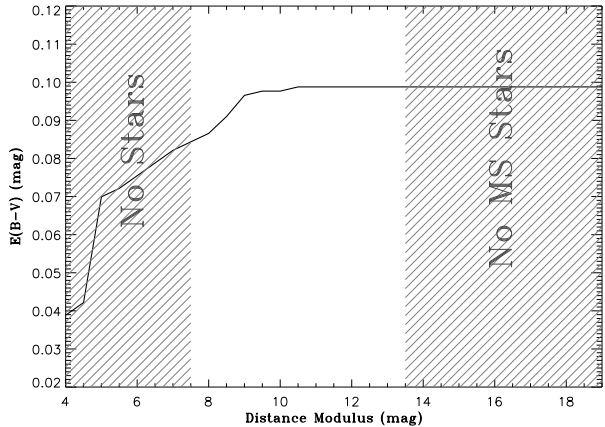}
\caption{Cumulative reddening as a function of distance modulus.}
\label{fig:dist_ebv}
\end{figure}

\cite{Green-15} derived a Galactic 3D reddening map using high-quality stellar
photometry of stars from Pan-STARRS 1, the Two Micron All Sky Survey (2MASS),
and Gaia, for the sky with decl. $\gtrsim~30^\circ$. The map has a hybrid angular
resolution, ranging from $3\farcm4$ to $13\farcm7$, with a typical value of $6\farcm8$,
which is comparable to the size of the ring-like nebula (10$'$ in diameter). We
downloaded the relation between $E(g - r)$ and distance from the
website\footnote{http://argonaut.skymaps.info} in the direction of the nebula.
According to the extinction factor listed in Table 1 in \cite{Green-18}, we
convert the $E(g - r)$ to $E(B - V)$ and show it in Figure \ref{fig:dist_ebv}. From
this figure, we can see the maximum $E(B - V)$ in this direction is 0.099 mag,
which is consistent with the result derived from IRAS and H {\sc i} data.  For
distance modulus (DM) less than 7.5 or greater than 13.5, shown by the
line-filled region, there is no star or no main-sequence (MS) star being
available to derive the cumulative reddening profile. Therefore, the steep rise
of $E(B - V)$ around the DM of 4.75 ($\sim$90 pc) cannot be identified as the
jump as the nebula.  Around the DM of 8.5, there is a relatively rapid rise of
$E(B - V)$ ranging from 0.04 to 0.07 mag, with an aptitude of 0.03 mag, which might
be from the nebula. In the following analyses, we then take the distance to the
nebula as 500 pc. We should note that the distance derived by this method has
large uncertainty, as the increase of the $E(B - V)$ with the DM is not steep
enough and the possibility of a smaller DM ($<$ 7.5) still exists. Besides, if
the distance to the nebula is 500 pc, it means most of the extinction along
the line of sight is actually the foreground extinction.

\section{Possible Formation Mechanism and Illumination Source}
\subsection{Star Forming Region}

The strong UV photons emitted by OB stars in the star forming regions can ionize
the surrounding ISM and the strong stellar wind can sweep away
the surrounding gases to form Str\"omgren sphere structures. However, the formation of
new stars requires the gas density to be high enough to collapse, hence the extinction
in the star formation region is usually high. For the nebula in this work, the
optical depth is quite low ($A_V\sim0.09$ mag), and there is no OB star found
in this region. So, this ring-shaped structure is not a Str\"omgren sphere
structure formed by star formation activities and the illumination source is
not OB stars.

\begin{figure}%[ht]
\centering
\includegraphics[width=\linewidth]{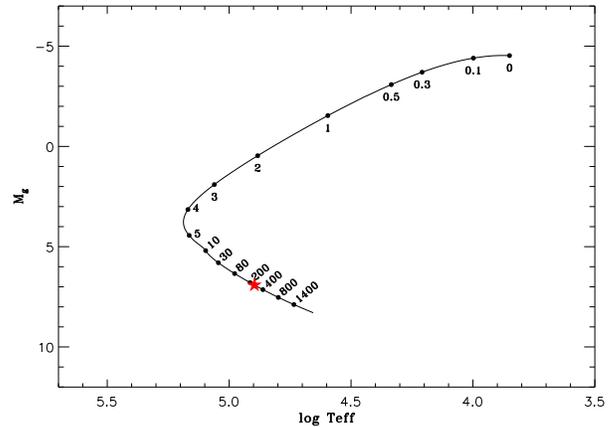}
\caption{Evolution of post-AGB H-burning model for the final mass of 0.576 $M_\odot$. The dot marks are in units of $10^3$ yr. The red star stands for the current position on the evolutionary track for the possible missing central star.}
\label{fig:Teff_Mg}
\end{figure}

\subsection{Supernova Remnant}

The morphologies of supernova remnants are complex and varied. In the new
catalog of 294 Galactic supernova remnants, 80\% of remnants show shell (or
possible shell) structures \citep{Green-19}. A shell can be formed when the
shock wave from the explosion sweeps the surrounding ISM, hence a ring-like
structure can be seen because of limb brightening. The size of a supernova
remnant is less than 1 pc in the phase of free expansion, and can reach to
more than 30 pc in the final phase of fading and death. As the physical size
of the nebula is only $1.74$ pc in diameter, and the velocity of the gas in
the ring is very small (see \S2.4), the possibility of the ring-shaped nebula
being a supernova remnant is ruled out.

\subsection{Planetary Nebula}

As mentioned in \S2.5, the velocity dispersion of H {\sc i} gas of the nebula is
1.02 $\rm km~s^{-1}$, which is significantly lower than the typical expanding
velocity (25 $\rm km~s^{-1}$) of a PN \citep{Kohoutek-01}. Hence the nebula might
be the relic of a PN. The inner and outer radii of the shell are $2\farcm3$ and
$5\farcm0$, respectively. Assuming the expansion velocity is 25 $\rm km~s^{-1}$, the
shell expanded to the median radius of the ring ($3\farcm65$, physical distance of
0.64 pc) from the center after 25,000 yr, then the shell became wider and
wider (physical thickness of 0.47 pc) because of the random dissipation (1.02
$\rm km s^{-1}$) over 230,000 yr, the current age of the nebula can be estimated
as about 255,000 yr. We should note that the calculation is very rough, as
(1) the expanding velocity of the shell depends on both the chemical of the
ejected material and the stellar core-mass evolution
\citep{Buzzoni-Arnaboldi-Corradi-06}, and (2) the mean velocity dispersion of the
shell should be higher than 1.02 $\rm km~s^{-1}$. 

With the help of theoretical evolution models of the post-AGB stars, we can
calculate the current absolute magnitude of the central star. From Figure
\ref{fig:Teff_Mg}, we show an evolutionary track of a post-AGB star with a
final stellar mass of 0.576 $M_\odot$, the dot marks are shown in units of
$10^3$ yr. The original data come from \cite{MillerBertolami-16}.
We translate the luminosity into the absolute magnitude in the $\textit{g}$ band by
assuming a blackbody radiation distribution for the central star.  The current
absolute magnitude and effective temperature of the central star are
$M_\textit{g}=6.91$ mag and log $T_{eff}$[K] = 4.90 respectively, about 255,000
yr after the star evolved into post-AGB stage.  Hence, there should be a
central white dwarf with a visual magnitude $\textit{g}=15.40$, which is much
brighter than the detection limit of the BASS survey. So the central star would
run away from this nebula if the system is the relic of a PN.

On the other hand, if the nebula is the relic of a PN, the gas in the ring can
be considered as the material ejected from the AGB envelope, hence we calculate
the mass of the ejected material by summing the H {\sc i} density in the ring
region. Given the distance of 500 pc, and the inner/outer radii of 0.40/0.87
pc, the H {\sc i} mass in the shell is estimated as $\rm 1.05 M_\odot$.

\begin{figure*}%[ht] 
\centering \includegraphics[width=\textwidth]{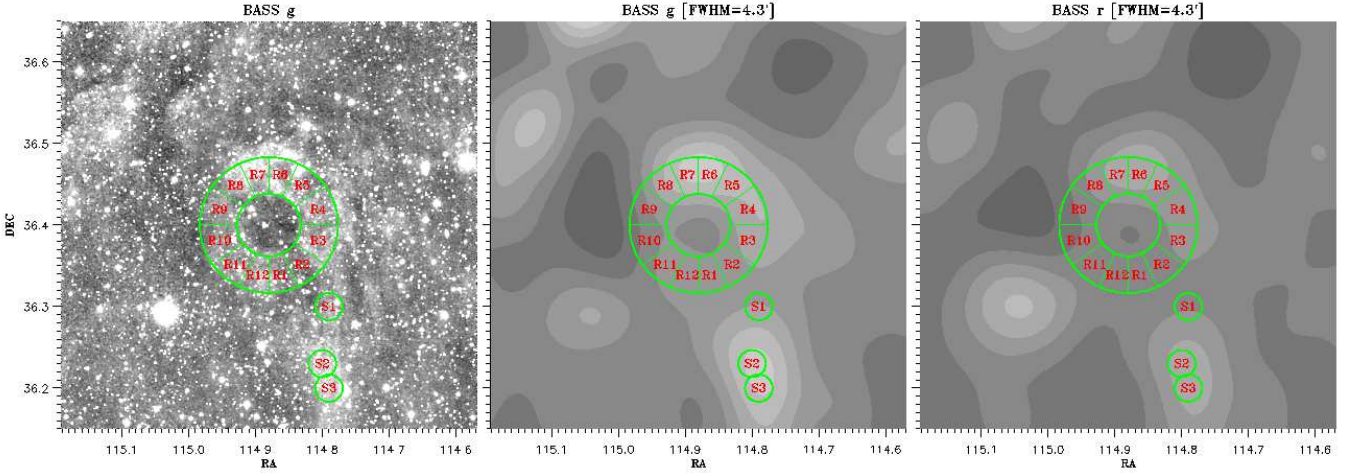}
\caption{The ring is divided into 12 regions according to the position angle to
the center, named as R1-R12. Three regions in the spur are also
selected and named S1-S3. The background is BASS images with a FOV of
$30'\times30'$. The BASS $\textit{g}$ image showed in Figure
\ref{fig:12bands} is replotted in the left panel but with a larger FOV. The BASS
$\textit{g}$ and $\textit{r}$ images are smoothed down to the resolution of the
IRAS 100 $\mu$m image, and are shown in the middle and right panels.}
\label{fig:15regions} 
\end{figure*}

Following \cite{Hildebrand-83} and \cite{Emsellem-95}, and assuming
average grain size of 0.1 $\mu$m,  the mass of dust in the ring can be
estimated from the IRAS flux: 
\begin{equation}
M_d=5.1\times10^{-11}I^c_{\nu}d^2\lambda^4[exp(\frac{1.44\times10^4}{\lambda_{\nu}T_d})-1]
\end{equation} 
where $I^c_{\nu}$ is the IRAS flux in mJy at wavelength
$\lambda_{\nu}$, $d$ is the distance in Mpc, and $T_d$ is the temperature
of the dust in K. The result gives an  $M_d$ of about 0.01 $M_\odot$. Hence the dust mass
relative to the hydrogen mass is about 0.0095, which is consistent with the value
of the Kramers-Kroing argument ($\gtrsim$0.0083, see Eq. 21.19 in
\cite{Draine-11}).

The ejected mass varies from 0.01 to 3 $M_\odot$, from differentiated PNe
from lower-mass nova shells to higher-mass ejecta shells from massive
Population I stars \citep{Frew-08}. So the total mass (H {\sc i} $+$ dust) of
$\rm 1.06 M_\odot$ in the shell is consistent with the assumption that the ring is
the material ejected from the envelope of the post-AGB stage. According to the
analyses shown above, this nebula is probably the relic of an evolved PN.

\subsection{Illumination Source}

At high Galactic latitude, most of the neutral gases with low velocities are in
a high degree of turbulence with gas temperatures of about a few thousand
Kelvin, hence their H {\sc i} 21 cm emission lines will show broad structures
\citep{Kalberla-Kerp-09}. On the other hand, narrow structures are rarely
observed at high latitudes and are associated with the infrared cirrus
\citep{Low-84}. As the infrared cirrus was explained as the reflection and
scattering of the DGL by the ISM, we then calculate the surface brightness of
the nebula to check whether the illumination energy mainly comes from the flux
of the total Galactic plane.

Following the work of \cite{Sandage-76}, the flux at any height $h$ above the
Galactic plane is equivalent to $m_V=-6.73$ visual magnitude, independent of
$h$. The surface brightness in the $\textit{V}$ band can be estimated by
$SB_V=22.7-2.5log(\gamma A_V / Q)$, where $\gamma$ is the albedo and $Q$ is the
efficiency factor of the grains. Given $\gamma=0.5$, $Q=2$, and $A_V=0.07$, the
$\rm SB_V$ is 27.1 $\rm mag~arcsec^{-2}$, indicating the DGL can provide sufficient
energy to illuminate the ring-like nebula.

As the nebula is illuminated by FUV photons and re-emitted in the longer
wavelength, the intensities of FUV, FIR, and H {\sc i} column density are expected
correlated with each other. The results of $N_{HI}=\rm9.97\times10^{19}~cm^{-2}$,
$\rm S^c_{100}=1.64~MJy~sr^{-1}$ (or $\rm 0.49\times10^{-10}~erg~s^{-1} cm^{-2}$
$\rm \AA^{-1}~sr^{-1}$) and $\rm S^c_{FUV}=0.74\times10^{-8}~erg~s^{-1}
cm^{-2}$ $\rm \AA^{-1}~sr^{-1}$ reasonably follow the linear relations shown
in Figure 2 and Figure 3 in \cite{Jakobsen-deVries-Paresce-87}.
Besides, the color $\textit{g-r}$=0.3 and $\rm S^c_{100}=1.64~MJy~sr^{-1}$
also follow the linear relation shown in Figure 10 in \cite{Roman-Trujillo-Montes-19}.

\begin{figure}%[ht]
\centering
\includegraphics[width=\linewidth]{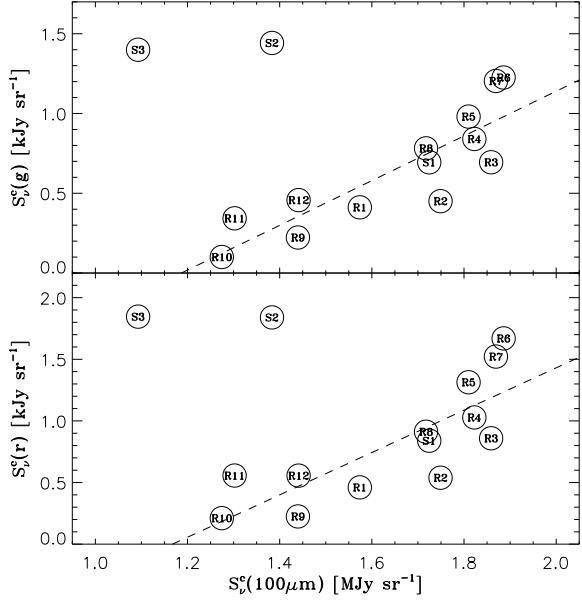}
\caption{Correlation of optical brightness against 100 $\mu$m brightness. The dashed line shows the result of the linear fitting of the regions of R1-R12.}
\label{fig:optical_ir}
\end{figure}

\begin{figure}%[ht]
\centering
\includegraphics[width=\linewidth]{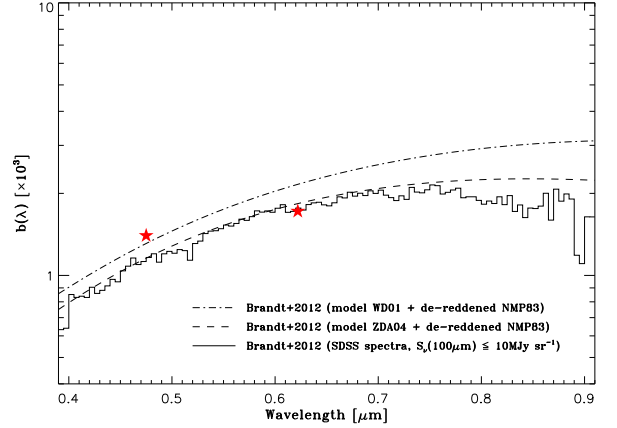}
\caption{Correlation slopes b{$_\lambda$} as a function of wavelength. The red stars are the slopes at $\textit{g}$ and $\textit{r}$ bands.}
\label{fig:slope}
\end{figure}

In Figure \ref{fig:15regions}, we divide the ring into 12 regions according to
the position angles to the center, and name them as R1-R12.  Three regions in
the spur, each with a radius of 1$^{\prime}$ and named S1-S3, are also
selected to make a comparison. Two different Gaussian smooth kernels are
applied to BASS $\textit{g}$ and $\textit{r}$ images, ensuring the optical data
match the resolution of the IRAS 100 $\mu$m data. The optical brightness as a
function of 100 $\mu$m is shown in Figure \ref{fig:optical_ir}. We fit the data
of R1-R12 by $S^c(\lambda) = a(\lambda) + b(\lambda) \times S^c_\nu(100\mu)$,
and get the correlation slopes $b(\textit{g})=1.40\times10^{-3}$ and
$b(\textit{r})=1.72\times10^{-3}$. We can see the region S1 follows the main
trend of the regions R1-R12, while S2 and S3 are obviously apart from the
trend, indicating S2 and S3 have different origins from S1 and the ring. In
Figure \ref{fig:slope}, we compare this results with the spectra of the DGL,
which were shown in Figure 6 in \cite{Ienaka-13}. The solid line is obtained by
analyzing the optical blank-sky spectrum from the SDSS by
\cite{Brandt-Draine-12}, while the dashed line and dashed-dotted line are model
spectra, based on the work of \cite{Zubko-Dwek-Arendt-04} and
\cite{Weingartner-Draine-01}, respectively. It should be noted that  (1)
the spectra showed in \cite{Ienaka-13} have been scaled with a biased factor of
2.1, and we find a lower biased factor of 1.62 is better to match our results, and
(2) the spectra shown here are weighted spectrum at intermediate Galactic
latitudes $|b|$ = 30$^\circ \sim 45^\circ$.  But as shown in Figure 4 in
\cite{Brandt-Draine-12}, the spectra change little with the selection of the
coverage of $|b|$, when $|b|$ is lower than $50^\circ$. We can see our results
are reasonably reproduced by the observational and model spectra of the DGL.

In summary, the analysis listed above supports the assumption that the
illumination source is the DGL. The ring structure clearly seen in optical
bands can be explained as the optical counterpart of the infrared cirrus.

\section{Conclusion and Discussion}

A ring-like nebula is found by BASS, with the $\textit{g}$ band surface
magnitude down to 27.42 $\rm mag~arcsec^{-2}$. Except for the $\textit{z}$ band from BASS
and the W1 and W2 bands from WISE, positive detections have been shown in multiband images
from FUV to FIR.  The properties of the nebula and main
conclusions can be summarized as follows.

\begin{itemize}

\item The H {\sc i} gas in the nebula has a very low velocity dispersion
($\rm \sigma_v \sim 1.02\ km\ s^{-1}$), and the dust is cold with a temperature of
$T_d\sim22$ K. The reddening of this nebula $E(B - V) \sim$ 0.02 mag is estimated
from IRAS 100 $\mu$m brightness and H {\sc i} column density. 

\item With the help of the 3D reddening map from Pan-STARRS 1, 2MASS, and Gaia,
the distance to the nebula from Earth is derived as 500 pc. The inner and the
outer regions are about 0.80 and 1.74 pc in diameter, respectively. The masses
of H {\sc i} and dust in the ring are measured as 1.05 $M_\odot$ and 0.01
$M_\odot$, respectively.  

\item Such a low-surface-brightness nebula whose energy can be
interpreted by the DGL could account for the optical counterpart of the
infrared cirrus which was detected by IRAS more than 30 yr ago.

\item The ring-like shape is not formed by recent star formation or supernova
explosion. It might be the ultimate phase of a PN, while the
central white dwarf star has been kicked off from the nebula for an unclear
reason.  

\end{itemize}

A spur is located to the south of the ring, which makes the overall structure look like the number
9. Is there a possible physical relation between the spur and the
ring? If the central star runs away from the nebula, it might distort the ring
shape \citep{Wareing-Zijlstra-O'Brien-07}. From Figure \ref{fig:optical_ir} we
can see S2 and S3 have quite different distributions in the optical brightness
versus the 100 $\mu$m brightness diagram with R1-R12, which indicates S2
and S3 have different origins from the ring. On the other hand, S2 follows the
main trend in the diagram, and the R2 region has a relatively lower brightness and
deviates from the trend. The nebula is too faint to be detected by normal
spectroscopic telescopes, so it is hard to obtain the spectra to make an actual
classification.  A long duration exposure of narrowband imaging would be a better
choice. So we are planning to observe this nebula using a H$\alpha$ or [S {\sc ii}]
filter to check whether the nebula is ionized.

There is large uncertainty about the distance estimation. For this line of sight, the
distance of 500 pc implies that the nebula is located at the edge of the dust
disk of the Milky Way. So the possibility of the nebula being typical ISM cannot
be ruled out. As the H {\sc i} signal is centrally distributed around
$\rm v_{lsr}\sim8.36~km~s^{-1}$, and the residual background signal around
$\rm v_{lsr}\sim0~km~s^{-1}$ is weaker, the possibility of them being chance
configurations in the line of sight is small. However, the ring
structure being a superposition of two close filaments  might be another
reasonable explanation. Besides, as the radial velocity of the nebula is
small, and the multiband data can be interpreted by the DGL, the nebula
should be a Galactic cirrus, rather than an extragalactic source.

Of the 210 PNe within 2.0 kpc, 90\% of PNe have identified central stars.
Most of the missing central stars are in more distant, reddened PNe, though
high internal extinction can also hide the central star \citep{Frew-08}.
Apparently, the nebula in this study does not belong to this case, because it
is near to us and the extinction along the light is quite low. 

There are some examples of other mechanisms to cause the loss of the central
stars of PNe. SBW 2, in the direction of the boundaries of the star-forming
Carina Nebula, has an ionized ring. However, no central star is found near its
center. Assuming pure hydrogen gas and fully ionization, the ring's mass is
estimated as 0.1 $M_\odot$. \cite{Smith-Bally-Walawender-07} have suggested
that the lack of any bright central star in SBW 2 may indicate that a binary
system has been disrupted. 

SuWt 2, in the direction of the constellation Centaurus, also has an ionized
ring and has lost the central white dwarf star. Instead, there is a 4.9 day period
binary consisting of two A-type MS stars, which are not hot enough
to make the nebula glow \citep{Bond-02}. It has been suggested that the stars
at the center of SuWt 2 were born as a family of three, with the A stars
circling each other tightly and a third star whose initial mass was greater
than 2.7 $M_\odot$ orbiting them around further out. The common envelope was ejected
into the orbit plane, generating the ring-shaped nebula seen today.  Eventually,
the remnant of the third star cooled and faded rapidly below the detection limit
of the International Ultraviolet Explorer \citep{Exter-10}. However, \cite{Jones-Boffin-17} concluded that the
A-type binary is merely a field star system, by chance lying in the same line
of sight of the nebula, and that it bears no relation to SuWt 2 or its
unidentified central star.

It is expected that more and more dark rings without central stars will be
found in the deep imaging surveys. The statistical analyses of such dark rings
will help us to understand the formation of the structure of the Galactic ISM and
the formation and evolution of PNe.  

\acknowledgments

The authors thank the anonymous referee for helpful comments that improved this
manuscript.  W.Z. thanks Chao Liu, Zheng Zheng, Yougang Wang and Jingkun Zhao
for helpful suggestions. F.Y. thanks Marko Krco for help on using GALFA-H {\sc i}
data cubes. This work is supported by the National Natural Science Foundation
of China (NSFC) (No. 11733006) and the Joint Research Fund in Astronomy (No.
U1531118) under a cooperative agreement between the NSFC and Chinese Academy of
Sciences (CAS). This work is also sponsored by a National Key R\&D Program
of China grant (No.  2017YFA0402704). 

BASS is a collaborative program between the National Astronomical Observatories
of the Chinese Academy of Science and Steward Observatory of the University of
Arizona. It is a key project of the Telescope Access Program (TAP), which has
been funded by the National Astronomical Observatories of China, the Chinese
Academy of Sciences (the Strategic Priority Research Program ``The Emergence of
Cosmological Structures" grant No. XDB09000000), and the Special Fund for
Astronomy from the Ministry of Finance. BASS is also supported by the External
Cooperation Program of the Chinese Academy of Sciences (grant No.
114A11KYSB20160057) and the Chinese National Natural Science Foundation (grant
No. 11433005). The BASS data release is based on the Chinese Virtual
Observatory (China-VO).

\bibliography{ref}

\label{lastpage}
\end{document}